 \documentclass[final,authoryear,3p,12pt,times]{elsarticle}
 \usepackage{graphics}
\usepackage{graphicx} 
\usepackage{amssymb}

\begin{document}

\begin{frontmatter}

\title{A Holistic Look at Requirements Engineering in the Gaming Industry}

 \author[label1]{Aftab Hussain}
  \author[label2]{Omar Asadi}
  \author[label3]{Debra J. Richardson}
 \address{Department of Informatics,\\ Donald Bren School of Information and Computer Sciences,\\ University of California, Irvine}
 \address[label1]{aftabh@uci.edu}
 \address[label2]{oasadi@uci.edu}
  \address[label3]{djr@uci.edu}

\begin{abstract}
In this work we present an account of the status of requirements engineering in the gaming industry. Recent papers in the area were surveyed. Characterizations of the gaming industry were deliberated upon by portraying its relations with the market industry. Some research directions in the area of requirements engineering in the gaming industry were also mentioned. 

\end{abstract}


\end{frontmatter}


\section{Introduction}
\label{sec:Intro}

The main goal of this work was to meticulously explore the status of requirements engineering in the gaming industry. There have been several market research studies illustrating the vast growth of the gaming industry. One such example is mentioned in~\cite{mobile}, where global revenues in the gaming industry was found to \$3.1 bn in as early as 2006. The market has grown significantly since then, making the industry hardly ignorable in terms of the economic value it generates. 

In this work, we observe the gaming industry with an requirements engineering perspective. In particular, we see how requirements engineering practices in the gaming industry differ from those in the traditional software industry. In order to do that we take a step by step approach of understanding the gaming industry. We first see how the gaming industry shares all the characteristics of any market driven industry (Section 2), where we then see the status of requirements engineering in the market driven software industry based on surveyed literature. In the next step we surveyed the general software engineering challenges that exist in game development, and how they can impact requirements engineering processes (Section 3). We then go deeper into studying the characteristics of game requirements and how some researchers have tried to use them to leverage the requirements engineering process (Section 4). In Section 5, we enlist current development trends in the gaming industry based on a number of surveyed papers, with a requirements engineering perspective. We present some future directions for research in the area of requirements engineering for games in Section 6, and conclude the paper in Section 7.

\section{An Overview of Software Development in the Market Driven Industry}
\label{sec:MrktDriven}

In this section, we emphasize upon software development in the market driven industry, with a focus on requirements engineering (RE from here onwards) primarily based on the findings of the work of~\cite{mrkt} et al. We first briefly explain how the games development industry is a market-driven industry, and how this industry differs from the traditional software development industry with respect to its target customers (Subsection~\ref{subsec:mrktChar}). Next, we see how the development process, in particular the RE phase, in the traditional industry compares with that in the market driven industry.

\subsection{The Gaming Industry: A Market Driven Industry}
\label{subsec:mrktChar}  

Among the most widely cited works on the characteristics of a market driven industry is by~\cite{market-drivenCore}. The following descriptions are given for a market driven industry in that work.

\begin{itemize}
\item ``Make the customer the final arbiter"
\item ``Understand our markets"
\item ``Commit to leadership in the markets we choose
to serve"
\item ``Deliver excellence in execution across our
enterprise."
\end{itemize}

From the above we can see that the current state of the market will drive the design decisions of the product. The seeds of design therefore typically start to grow from understanding the characteristics of popular trends within the domain of the product, who are the main competitors in the domain, what are the characteristics of the target market segment, etc. The findings will then set the tone for the development of the entire product.

The most defining property of a market-driven industry is that it targets the mass market, where even a small segment of the market can include a large number of diverse customers (~\cite{mobile}). This is typically not the case for the traditional software engineering industry, where customers are generally known to the software vendors and projects are customer-specific. The projects are built ground-up in a way where the goal is to meet the goals of the specific customer rather than to abide or follow the domain trends. Also in those industries, newer projects are often improvised versions of previous projects, where previous designs are strictly observed. The design development process thus typically follows a top-down approach.       
  
Based on the above mentioned reasons, it is justifiable to classify the gaming industry as a market-driven industry. It is geared towards serving a large spectrum of customers. Since identifying specific evolving needs of each customer is challenging, the corporations in the games industry pay a lot of emphasis on studying market research data, game postmortem websites and game review websites like Metacritic and PCgamer to understand the recent trends in the gaming industry. For instance, a number of recent (as of 2013) trends are mentioned by~\cite{5trends}; live-streaming and video sharing will become standard for games and gaming platforms, virtual reality capabilities of games, console providers like Sony, Microsoft, and Nintendo opening up to independent developers to publish games on the platforms of the console providers, offering games before completion because players now are willing to experience a game before they are published, etc. These trends originate from the market and are crucial to observe while designing games in order to compete in the games sector. The market driven nature of games development thus cannot be over emphasized. 

In the next subsection, we see how development practices, in particular RE, are performed in a market driven industry.  

\subsection{RE in the Market Driven Software Industry}
\label{subsec:REmrkt} 

For understanding the characteristics of RE practices in the market driven industry, we shall mainly draw upon the findings of~\cite{mrkt} (\ref{subsubsec:REVar}). But first we shall briefly explain the different stages of RE in a traditional software engineering setting as per the definitions of~\cite{Cheng:2007} (\ref{subsubsec:reprelim}). 

\subsubsection{Generic Phases in RE (\cite{Cheng:2007})}
\label{subsubsec:reprelim}

\textit{Elicitation}. This phase primarily comprises of understanding the goals of the system and delineating the requirements that need to be met in order to fulfill those goals. The requirements may include devising new functionalities, modifying old properties of a system, etc. Techniques used for this phase typically include stakeholder analysis (to ensure information is elicited from all the relevant stakeholders), brainstorming techniques for coming up with new requirements, and feedback techniques for increasing the accuracy of the requirements. These processes may take the help of models for guidance. As elicitation is an early phase of RE, the models used at this stage may not necessarily be accurate or highly formalized.

\textit{Modelling}. This phase begins when the requirements have been refined to an acceptable level of accuracy. Modelling helps to codify the requirements in a consistent, coherent, and structured fashion with the help of formal notations. Formal notations raise the level of abstraction and specify detailed requirements of the system, such as the kind of data that is stored by an application, and the kinds of operations that are performed on the data. The models generated are thus less likely to be misunderstood by workers, for instance developers, in the subsequent phases of the development process. Modelling is a challenging phase as it entails, optimally identifying the different subgroups of stakeholders/concerns/functionalities of a system and looking for opportunities for transforming or combining models, etc. Different modelling techniques are in use in industry. One such example is scenario-based modelling. 

\textit{Requirements Analysis}. This phase comprises of analyzing the correctness of the requirements, i.e., looking for possibly conflicting requirements, under-defined requirements, missed assumptions, etc. The phase might require going back to elicitation for further refinement of the requirements. Different techniques have emerged to identify an optimal requirements set.  

\textit{Validation and Verification}. Validation is by and large an informal process of checking whether the requirements in the requirements document of a project indeed satisfies the objectives of the stakeholders. Verification is a more rigorous activity as it involves proving whether specific requirements specifications provides the desired objectives. That is why verification is only possible when a formal model of requirements exists.  

\textit{Requirements Management}. This phase involves tracing requirements artifacts and connecting them with various software artifacts further down the development stream, identifying opportunities for requirements evolution, getting in analytics and visualizations for requirements in order to aid future project decisions.  

\subsubsection{Variation in RE in the Market Driven Software Industry (\cite{mrkt})}
\label{subsubsec:REVar}

Dahlstedt et al. performed a literature survey on market driven RE. They found that requirements of low priority are shelved in order to facilitate the early release of products, which is a critical aspect in market driven software companies. Also, the requirements are initially invented by the development organization rather than elicited from the customers, since there is no well-specified customer segment. It is only later that customer inputs are incorporated with the help of feedback mechanisms. This lack of specificity in the description of customers could also be a reason behind not having significant requirements documentation or modelling. Verbal communication is the primary means for conveying requirements. Validation tends to take the back seat in RE in the market-driven setting as well and this could also be attributed to the absence of a specific customer. Some forms of validation are generally post-release (through market surveys, prototype reviews, etc). Requirements management, however, has been identified as an important area of market driven development because of the crucial benefit it could offer. Having seamless requirements tracing mechanisms would enhance tracking product evolution and even maintenance. The use of a common repository for storing the requirements has been advocated for this purpose.

Two distinguishing RE phases in the market driven setting is requirements prioritisation and release planning. As time-to-release is a survival attribute in market driven development, it becomes important to sieve out the most important requirements and also check for any interdependencies between requirements. 

Dahlstedt et al. also performed a qualitative research study using semi-structured interviews with seven Swedish software companies with a market-driven development focus in order to investigate to what degree their literature survey findings were practically observed at the time. Although most of those findings matched with their practical findings, there were a few differences. For instance, time-to-release was more of a crucial attribute for those companies who had more competitors in their domain. Also, all their surveyed companies did document requirements.  

\section{Challenges in Game Development}
\label{sec:GameDevProblems}

In this section, we shall visit the challenges in game development and how some of them pose a problem to RE. Because project data are hard to obtain from game companies\footnote{Game companies do not reveal data in order to maintain a competitive edge with their domain counterparts. (\cite{Petrillo:2009:WWS:1486508.1486521})}, most previous works [\cite{flynt}, \cite{Petrillo:2009:WWS:1486508.1486521}, \cite{bethke}] have reviewed publicly available game postmortems to elicit the characteristics and challenges in game development. We first briefly explain postmortems in Subsection~\ref{subsec:post-mortem}, and then discuss problems found in Subsection~\ref{subsec:gamesProblems}.

\subsection{Postmortems}
\label{subsec:post-mortem}

Game postmortems are reviews of completed game projects written by developers who have actually worked on them. In these review developers write about the experience they had while building their games: the software development methodology they used, the issues they faced on each phase of the software development, issues they encountered with project coordination, etc. Postmortems are posted on popular game review sites like Gamasutra, where other registered users of the site can weigh in their opinions on the postmortems as well. Postmortems do not have any fixed structure, and are written in a very informal, note-taking style. However, they create fruitful discussion seeds on game development issues. Below is an extract from a postmortem of the game Diablo 2 developed by Blizzard Entertainment\footnote{http://www.gamasutra.com/view/feature/131533/postmortem\_blizzards\_diablo\_ii.php (accessed March 2015)}.

\textit{``The Diablo II team comprised three main groups: programming, character art (everything that moves), and background art (everything that doesn't move), with roughly a dozen members each. Design was a largely open process, with members of all teams contributing. Blizzard Irvine helped out with network code and Battle.net support. The Blizzard film department (also in Irvine) contributed the cinematic sequences that bracket each of Diablo's acts, and collaborated on the story line."}

Despite their informalism, postmortems reveal current practices that are being adopted for building games in the real world, which could assist in building perspectives on the development side of games. 

\subsection{Problems}
\label{subsec:gamesProblems}

Based on their literature study, \cite{Petrillo:2009:WWS:1486508.1486521} elucidate a number of problems that are encountered in game development: (1) scope problems, for instance, feature creep - initial scope of the project is inadequately defined and consequent development entails addition of new functionalities which becomes an issue for requirements management, (2) scheduling problems - underestimating the times various project tasks could take and consequently setting ambitious deadlines for milestones that are hard to meet, (3) crunch times, i.e., times before finalization of a project tend to get more work heavy in the gaming industry than crunch times in the software industry, (4) technological problems that may arise due to working on new platforms that have not sufficiently matured with respect to deployment, etc. It ought to be noted that in addition to these problems, common problems in the traditional software industry are also found in the gaming industry (\cite{Petrillo:2009:WWS:1486508.1486521}). 

\begin{figure}[!htbp]
\centering
\includegraphics[width=6in]{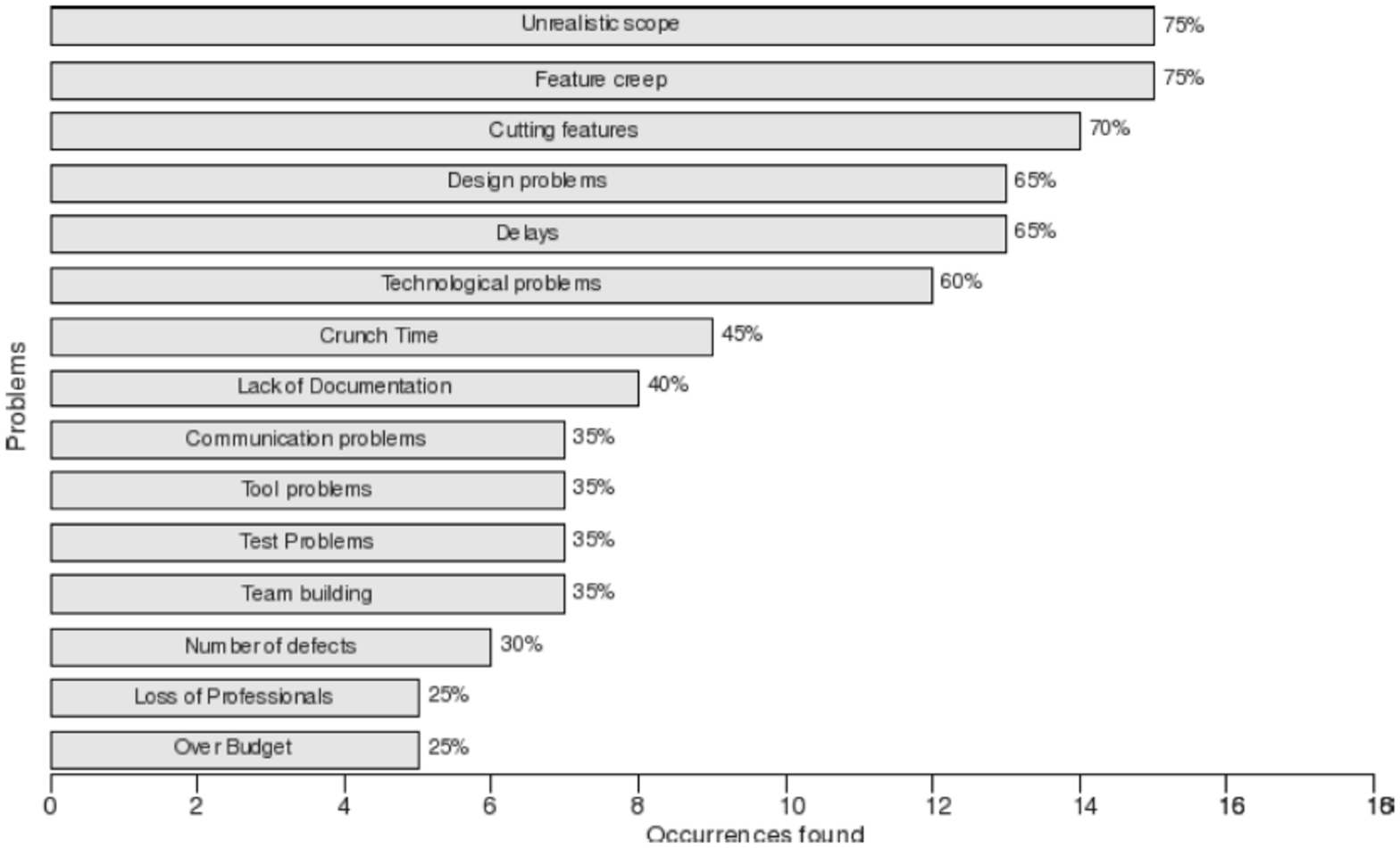}
\caption{Common Problems in Games Development.~\cite{Petrillo:2009:WWS:1486508.1486521}}
\label{fig:GamesProbs}
\end{figure}

\cite{Petrillo:2009:WWS:1486508.1486521} also performed a survey of 20 postmortems of completed game projects to elicit what software development problems were practically facing within the gaming industry. The most common problems found are showed in Figure~\ref{fig:GamesProbs} which was obtained from their paper. Problems of ``over budget" and ``crunch time", although mentioned to be prevalent in literature, were found to be less significant problems in the projects they studied by Petrillo et al.

Overall, most problems of the gaming industry are shared with those of the traditional software industry. However, there are a few differences. For instance, game development teams consist of more diverse teams in terms of job roles - there will be story writers, UX designers, developers, music artists, etc. This could lead to problematic issues for communication as each have different perspectives - this could lead to splits within the team (\cite{Petrillo:2009:WWS:1486508.1486521}). In addition, elaboration of games requirements is tricky as most of them are related to the ``fun factor" which is highly subjective in nature (\cite{Petrillo:2009:WWS:1486508.1486521}). In the next section, we shall take a look at some works that have attempted to logically navigate such requirements in games. 

\section{Understanding Games Requirements}
\label{sec:gamesReq}

The previous sections have elaborated on the nature of the gaming industry and thus formed a foundation for understanding the impact on RE in the gaming industry. In particular, we can now understand that(~\cite{mobile}): (1) elicitation of requirements is difficult from the broad customer set that typical gaming companies have, (2) the subjective nature of the requirements further complicate requirements elicitation, (3) being a highly creative industry, the gaming industry will always find it challenging to balance out ``fun factor" requirements with critical non-functional requirements(\cite{Callele:Creative}). In this section, we look at Callele's works on understanding game-specific requirements. 

\subsection{Callele's Works on Game Requirements}

Satisfying emotional requirements are highly important with regard to the success of a game. The main challenge of working with emotional requirements with the design phase has been with accurately specifying and representing them due to their highly subjective nature. Callele et al. in \cite{Callele:emotionalreq} and a number of other works propose methods that take a considerable step in deciphering emotional requirements. In [\cite{Callele:emotionalreq}], they dichotomize emotional requirements into two aspects: intent and artistic context. The \textit{intent} of an emotional requirement is a verbalization of exactly what the designer of the game wants the player to feel in a certain moment of the game. The \textit{artistic context} is the means which the developer would use to actually induce those feelings in the gamer. 

Callele et al. also proposed visualizations for emotional requirements. For instance, they proposed emotional terrain maps, emotional intensity
maps, and emotion timelines as visual mechanisms.  

\cite{CalleleSecurityEmotional} presents a very interesting work in how emotional requirements could be used in prioritizing security requirements for the game. The main motivation of the work was that emotional discontentments in the game could override security requirements, and thus it is important to gauge the security risk with respect to the emotional dissatisfaction it can cause in among the consumers. Other works of Callele include the proposal of cognitive requirements for games (~\cite{CalleleCognitive}), and the embodiment of game-specific requirements into experience requirements (~\cite{CalleleExperienceReq}). 

\section{Current Development Practices in the Gaming Industry}
\label{sec:ReTechinGameIndustry}

In this section, we look at a few works in order to portray the general, recent, themes in software development within the gaming industry (Subsections~\ref{subsec:model} to \ref{subsec:deadline}). We take stock of these works with an RE perspective in Subsection~\ref{subsec:link}.

\subsection{Model-Driven Game Development (\cite{ModelDriven})}
\label{subsec:model}

	In this work, the authors discuss a model-driven approach to modern video game development. In the past, games could be developed by a single programmer in a few weeks' worth of work. Now, video games need teams of dozens to hundreds of people and several years of work. The teams include artists, marketers, voice actors, and musicians in addition to the programmers. As such, the development teams needed to move away from ad-hoc development and waterfall methodologies into something more suitable for the colossal task of creating such a complex software system. The teams tried incorporating component-based development and agile methodologies with some success. The paper seeks to extend this success to model-driven development of computer games.
	The main goal of model-driven development here is to abstract away the finer details of the game. This is done with high-level models such as structure and behavior diagrams as well as control diagrams. By expressing the game's design in this way, it allows management to gain at least a rudimentary understanding of the game design and help alleviate the communication problems encountered in previous development methodologies.

\subsection{Cheating: Gaining Advantage in Videogames (\cite{cheating} based on review of \cite{Review})}

	In this review, the author reviews the book Cheating: Gaining Advantage in Videogames by Mia Consalvo. He tackles the questions Consalvo proposed in the book, including who should define cheating and how it should be policed. The first part of Consalvo's book offers a historical journey through the culture of video games from the early days in 1980 to the present. The reviewer took note of the battle between manufacturers who wish to exert control over the players and external forces who try to profit by undermining the manufacturers. In the middle of this battle are the players who seek to make a name for themselves as top gamers.
	
	The second part of Consalvo's book narrows its focus to how players interact in online multiplayer games and how cheating impacts this form of play. Consalvo identified three types of players: purists, moderates, and hardcore players. Purists believe that players should not be seeking external aid other than asking friends for advice. Moderates believe that walk-throughs and guides are acceptable forms of aid. Hardcore players believe that only other players can be cheated, not the games themselves.
	
	The remainder of Consalvo's book focuses on the social interactions associated with cheating. Parts of these interactions include the search for cheating programs to gain unfair advantages over other players. Game developers should address these possibilities in the design of their games. Perhaps the anti-cheating measures can be captured at the model level as specified in the previous paper.

\subsection{Comparative Analysis of Porting Strategies in J2ME Games (\cite{Alves})}

	In this paper, the authors discuss the problem of porting mobile games from one device to another. The ever-increasing number of mobile devices combined with the ever-shortening release cycle for these mobile devices is putting a new type of pressure on mobile game developers to quickly port and release versions of their mobile games for all available mobile devices. Here, the authors present and discuss several of the challenges facing mobile game developers with regard to porting the mobile games from device to device. The challenges include different device hardware features, different execution availability and application size limits, different profiles, different implementations of the same profile, proprietary APIs, device-specific bugs, and internationalization.
	
	The authors then discussed various case studies of mobile games they ported from one device to another. In each case, a different porting strategy was used and analyzed. Each presented their own drawbacks, such as poor code maintainability or highly granular code changes necessary.
	The challenges that these authors presented with porting mobile games exist for desktop games as well. In the world of computer gaming, it's very likely that each gamer's computer has different hardware than any other computer. The developers must acknowledge this during the design phase so that the game can be developed in a way that can scale automatically up or down depending on the hardware.

\subsection{Proposal of Game Design Document from Software Engineering Requirements Perspective (\cite{Mitre})}

	In this paper, the authors propose a Game Design Document as an improvement over the existing standard for System Requirements Specifications. The document consists of several important sections, which will be outlined below. The authors’ goal with this document is to improve upon the understanding provided by a System Requirements Specification.
	
	The document’s first section is the overview. This section summarizes the objectives of the game and attempts to keep the team on-task. The next section is game mechanics, which deals with the game elements present. Next is the game dynamics, referring to any interactions within the game. Following that is the aesthetics section, dealing with anything the player perceives with their senses. The next section, experience, addresses the expected player experience. Finally, the assumptions and constraints list all the technical limitations present.
	
	These sections of the Game Design Document improve upon the System Requirements Specification by adding information that was not previously contained, such as the aesthetics of the system. This improvement provides a more complete specification of the game and allows for better communication between different levels within the development team.

\subsection{Extending Use Cases to Support Activity Design in Pervasive Mobile Games (\cite{ExtendingUseCases})}

	In this paper, the authors propose a language to define activities in mobile games. The authors point out that there has been very little research in this area and that there was a dire need of a bridge between the preproduction and production phases of development. There needed to be a way to design activities in these very diverse and complex processes. The main focus here is that traditionally, software has been focused on productivity, which is a measurable idea. Games, however, focus on entertainment, which is completely subjective and cannot be reliably measured. A secondary focus is that games are creative projects rather than strictly engineering projects, which adds a preproduction phase to development not present in traditional software. The gap between this new preproduction phase and traditional production can cause many games to fail before they have been completed, which is why the authors sought to introduce this new language to bridge the gap between the two phases.
	
	This paper attempts to tackle one of the non-functional requirements of games: enjoyability and entertainment value. Since these cannot be measured reliably, the only way to improve them is to streamline the development process rather than the product itself.

	This paper also attempts to tackle the root cause of nearly all defects in requirements: communication. By creating a new language with which developers can communicate, the authors seek to solve the problem of misunderstanding one another by ensuring a formal language about which there is no ambiguity whatsoever.

\subsection{Game Development: Harder Than You Think (\cite{Blow:2004:GDH:971564.971590})}

	In this work, the author discusses the reasons why game development today is so much harder than it was only 20 years ago. He divides the difficulties into two categories, difficulty from size and complexity and difficulty from highly domain-specific requirements, but notes that there is significant overlap between the two when the domain-specific requirements force the project to become more complex.
	
	The author identifies the tools, environments, and workflows as part of the reason for incredibly complex software projects in the gaming industry. According to the author, the current development environment on Windows was not meant for the type of development necessary for video games. Microsoft Visual Studio was built for Visual Basic and C\# mainly, not for C++, but the only viable C++ compiler and environment on Windows remains Visual C++. However, even games are developed to run on multiple platforms, and this means that the tools and environments need to be synchronized across them all. The author also identified that many companies do not consider the complexity of the code when they establish their workflows and do not allot any time for necessary code refactoring, contributing to the complexity of the code and increasing the amount of time it takes to develop quality code.
	
	The author identifies engine code, depth of simulation, and profiling as some of the highly domain-specific requirements that contribute to the difficulty of writing video games. The author further breaks down engine code into mathematical and algorithmic knowledge and the wisdom to know how the algorithms will interact when coupled together. The simulations in video games are incredibly complex themselves and require much more domain-specific knowledge than is normally required for software. Also, since the simulation is intended to run in real time, the developers would like to ensure that all resources are being used as efficiently as possible, leading to the profiling of CPU and GPU usage to determine where inefficiencies lie.

\subsection{Component Based Game Development – A Solution to Escalating Costs and Expanding Deadlines? (\cite{Folmer:2007:CBG:1770657.1770663})}
\label{subsec:deadline}

	Keeping in mind the ideas about complexity from the previous paper, this paper discusses the concept of using commercial off-the-shelf (COTS) components in video games rather than writing all the components from scratch. The author argues that development teams can significantly reduce development time in this way but then run the risk of struggling with component integration and architecture complexity. The paper also touches on some domain-specific components and their problems.

	Many of the decisions that must be made regarding using COTS components in a video game boil down to what the COTS components offer the developers. There is no point in using a COTS component if it does not perform a necessary function in the code.

\subsection{Linking it all together with the RE perspective}
\label{subsec:link}

	Each of these papers presents some isolated aspect of requirements engineering in video game development. Reyno and Cubel proposed using model-driven development to enhance communication between developers and management. Along the same lines, Valente and Feijó proposed a new formal language with which developers can clearly communicate with each other regarding intended activities in the video game. Again addressing communication among the development team, Cuauhtémoc and Sánchez proposed a more detailed Game Design Document that offers more information than a System Requirements Specification.
	
Consalvo identified that video game developers need to consider social interactions in their games while determining the requirements due to the prevalence of competition among players and some players' tendency to cheat the system to beat other players. These social actions are the defining characteristics of many video games today and must be accounted for in the development process.
Both Blow and Folmer addressed the rapidly increasing complexity problems of modern video game development and stressed that these issues must be tackled sooner rather than later. Blow mentioned the issues of workflow on a complex project while Folmer tackled the problem of COTS components increasing the complexity of the project.

\section{Research Opportunities for RE in Games}
\label{sec:ReIO}

In this section, we look into a number of opportunities for research in RE within the gaming industry based on the findings of \cite{Callele:rsrchOpp}(Subsection~\ref{subsec:oppcallele}), and suggest a relatively novel research area of requirements engineering by relating to a recent work that we did (Subsection~\ref{subsec:revrsEnggReq}). 

\subsection{Research Opportunities (\cite{Callele:rsrchOpp})}
\label{subsec:oppcallele}

Callele et al. cites that there is considerable research potential into the contextual interpretation of requirements. They give examples of how a certain requirement may be viewed as a functional requirement in the story design context but may be viewed as a non-functional requirement in the implementation phase. There is also work left to be done to view how the interpretation of requirements as functional or non-functional can have an impact on the different phases of RE, like elicitation, requirements validation, and requirements management.
Another interesting research opportunity elicited by Callele et al., which is based on their work on prioritizing security requirements with the help of emotional requirements, is how emotional requirements could be used to formulate requirements to identify and handle destructive stakeholders (for example, players who deliberately destroy other players in an action game) in the run of play. 
Finally there are research opportunities in understanding the reasons why certain game genres are popular in different segments of the consumer market - such research could begin investigations by dividing the market into segments based on age or geographic location. The findings from such explorations would help in predicting needs and thus building more quality requirements for games. 

\subsection{Reverse Engineering Requirements in Games}
\label{subsec:revrsEnggReq} 

There is significant scope in finding efficient and effective techniques in reverse engineering requirements in games. The development of the requirements document (or the Games Design Document) is not a standardized and thorough process in the games industry. This is mainly due to the diverse nature of the customers (As mentioned in Section~\ref{sec:MrktDriven}) that makes requirement elicitation challenging (~\cite{ElicitMgt}). We thus believe reverse engineering requirements from games can be a viable option to generate requirements, particularly for reuse of games design and components in the future. However, there has not been much work in this aspect. Reverse engineering descriptions of large legacy systems is challenging, particularly in eliciting in all non conflicting descriptions of all components of the game. There have been recommendations of reverse engineering requirements in an incremental way, i.e. decomposing the system into subsystems and generating requirements models for each of those subsystems~\cite{revReq}. 

In [\cite{Ourreport}], we reverse engineered the requirements of a popular  multiplayer, online simulation game, Euro Truck Simulator 2 built by SCS Software\footnote{The game has sold over 500,000 units as of April 2014, http://en.wikipedia.org/wiki/Euro\_Truck\_Simulator\_2}. We tried to observe the decompose-and-construct philosophy suggested in~\cite{revReq}. To this end, we found two techniques to be useful: the $i*$ approach~(\cite{Yu:1997:TMR:827255.827807}) to generate the system goal models and the rich picture method (\cite{Monk:1998:MAT:274430.274434}) to generate the system vision model. Our work, however, adopted an entirely manual approach. We believe there is considerable scope for research in seamlessly integrating reverse engineering techniques that are prevalent in architectural retrieval in the RE frame. In particular, there is scope for identifying requirements patterns, and exploring opportunities for automatically codifying requirements, perhaps with the help of formal notations. 

\section{Conclusion}
\label{sec:concl}

In this work, we have explored the status of requirements engineering in the gaming industry. We have studied how the gaming industry shares many of the traits of a market driven industry. In addition, we saw what problems are unique in the gaming industry, particularly with regard to their comparison with the traditional software industry. Literature pertaining to software development in the gaming industry and an assessment of how they impact RE process was made. We also reviewed some recent works on gauging requirements that are unique to the creative industry. We believe streamlining RE processes in the gaming industry can  help in resolving many of the bottlenecks that exist in the gaming industry. For reusing RE techniques in the traditional software industry may not be sufficient, particularly because of the fact that requirements in the gaming industry are distinctive in nature. Novel approaches, such those proposed in [~\cite{CalleleSecurityEmotional}] need to explored.

\bibliographystyle{elsarticle-harv}
\bibliography{biblio}

\begin{thebibliography}{27}
\expandafter\ifx\csname natexlab\endcsname\relax\def\natexlab#1{#1}\fi
\expandafter\ifx\csname url\endcsname\relax
  \def\url#1{\texttt{#1}}\fi
\expandafter\ifx\csname urlprefix\endcsname\relax\def\urlprefix{URL }\fi

\bibitem[{Alves et~al.(2007)Alves, Ramalho, and Damasceno}]{mobile}
Alves, C.~F., Ramalho, G., Damasceno, A. L.~G., 2007. Challenges in
  requirements engineering for mobile games development: The meantime case
  study. In: RE. IEEE, pp. 275--280.

\bibitem[{Alves et~al.(2005)Alves, Cardim, Vital, Sampaio, Damasceno, Borba,
  and Ramalho}]{Alves}
Alves, V., Cardim, I., Vital, H., Sampaio, P. H.~M., Damasceno, A. L.~G.,
  Borba, P., Ramalho, G., 2005. Comparative analysis of porting strategies in
  j2me games. In: ICSM. IEEE Computer Society, pp. 123--132.

\bibitem[{Asadi and Hussain(2015)}]{Ourreport}
Asadi, O., Hussain, A., 2015. Euro truck simulator 2: Reverse engineered
  requirements document. Tech. rep., Department of Informatics, Bren School of
  ICS, University of California, Irvine.

\bibitem[{Bethke(2003)}]{bethke}
Bethke, E., 2003. Game Development and Production. Wordware Publishing.

\bibitem[{Blow(2004)}]{Blow:2004:GDH:971564.971590}
Blow, J., Feb. 2004. Game development: Harder than you think. Queue 1~(10),
  28--37.
\newline\urlprefix\url{http://doi.acm.org/10.1145/971564.971590}

\bibitem[{Blueprint(2010)}]{revReq}
Blueprint, 2010. Reverse engineering requirements.
\newline\urlprefix\url{http://www.blueprintsys.com/back\_to\_the\_future\_reverse\_engineering\_requirements/}

\bibitem[{Callele et~al.(2005)Callele, Neufeld, and
  Schneider}]{Callele:Creative}
Callele, D., Neufeld, E., Schneider, K., 2005. Requirements engineering and the
  creative process in the video game industry. In: Proceedings of the 13th IEEE
  International Conference on Requirements Engineering. RE '05. IEEE Computer
  Society, Washington, DC, USA, pp. 240--252.
\newline\urlprefix\url{http://dx.doi.org/10.1109/RE.2005.58}

\bibitem[{Callele et~al.(2006)Callele, Neufeld, and
  Schneider}]{Callele:emotionalreq}
Callele, D., Neufeld, E., Schneider, K., 2006. Emotional requirements in video
  games. In: Proceedings of the 14th IEEE International Requirements
  Engineering Conference. RE '06. IEEE Computer Society, Washington, DC, USA,
  pp. 292--295.
\newline\urlprefix\url{http://dx.doi.org/10.1109/RE.2006.19}

\bibitem[{Callele et~al.(2008)Callele, Neufeld, and
  Schneider}]{CalleleSecurityEmotional}
Callele, D., Neufeld, E., Schneider, K., 2008. Balancing security requirements
  and emotional requirements in video games. In: Proceedings of the 2008 16th
  IEEE International Requirements Engineering Conference. pp. 319--320.

\bibitem[{Callele et~al.(2010{\natexlab{a}})Callele, Neufeld, and
  Schneider}]{CalleleExperienceReq}
Callele, D., Neufeld, E., Schneider, K., 2010{\natexlab{a}}. An introduction to
  experience requirements. In: Proceedings of the 2010 18th IEEE International
  Requirements Engineering Conference. pp. 395--396.

\bibitem[{Callele et~al.(2010{\natexlab{b}})Callele, Neufeld, and
  Schneider}]{CalleleCognitive}
Callele, D., Neufeld, E., Schneider, K., 2010{\natexlab{b}}. A proposal for
  cognitive gameplay requirements. In: Requirements Engineering Visualization
  (REV), 2010 Fifth International Workshop on. pp. 43--52.

\bibitem[{Callele et~al.(2011)Callele, Neufeld, and
  Schneider}]{Callele:rsrchOpp}
Callele, D., Neufeld, E., Schneider, K., 2011. A report on select research
  opportunities in requirements engineering for videogame development. In:
  Fourth International Workshop on Multimedia and Enjoyable Requirements
  Engineering {(MERE} 2011) - Beyond Mere Descriptions and with More Fun and
  Games , Trento, Italy, August 30, 2011. pp. 26--33.
\newline\urlprefix\url{http://dx.doi.org/10.1109/MERE.2011.6043942}

\bibitem[{Cheng and Atlee(2007)}]{Cheng:2007}
Cheng, B. H.~C., Atlee, J.~M., 2007. Research directions in requirements
  engineering. In: 2007 Future of Software Engineering. FOSE '07. IEEE Computer
  Society, Washington, DC, USA, pp. 285--303.
\newline\urlprefix\url{http://dx.doi.org/10.1109/FOSE.2007.17}

\bibitem[{Consalvo(2007)}]{cheating}
Consalvo, M., 2007. Cheating: Gaining Advantage in Videogames. The MIT Press,
  Cambridge, MA.

\bibitem[{Dahlstedt et~al.(2003)Dahlstedt, Karlsson, Persson1, och Dag, and
  Regnell}]{mrkt}
Dahlstedt, A.~G., Karlsson, L., Persson1, A., och Dag, J.~N., Regnell, B.,
  2003. Market-driven requirements engineering processes for software products
  - a report on current practices. Submitted to Development of Product Software
  14.

\bibitem[{Day(1998)}]{market-drivenCore}
Day, G.~S., 1998. What does it mean to be market-driven? Business Strategy
  Review 9~(1), 1--14.
\newline\urlprefix\url{http://dx.doi.org/10.1111/1467-8616.00051}

\bibitem[{Flynt and Salem.(2004)}]{flynt}
Flynt, J.~P., Salem., O., 2004. Software Engineering for Game Developers.
  Software Engineering Series. Course Technology.

\bibitem[{Folmer(2007)}]{Folmer:2007:CBG:1770657.1770663}
Folmer, E., 2007. Component based game development: A solution to escalating
  costs and expanding deadlines? In: Proceedings of the 10th International
  Conference on Component-based Software Engineering. CBSE'07. Springer-Verlag,
  Berlin, Heidelberg, pp. 66--73.
\newline\urlprefix\url{http://dl.acm.org/citation.cfm?id=1770657.1770663}

\bibitem[{Graft(2013)}]{5trends}
Graft, K., December 2013. The 5 trends that defined the game industry in 2013.
\newline\urlprefix\url{http://www.gamasutra.com/view/news/207021/The\_5\_trends\_that\_defined\\\_the\_game\_industry\_in\_2013.php}

\bibitem[{Henricks(2009)}]{Review}
Henricks, T.~S., 2009. Research directions in requirements engineering. In:
  American Journal of Play. pp. 532--534.
\newline\urlprefix\url{http://www.journalofplay.org/sites/www.journalofplay.org/files/pdf-articles/
  1-4-book-review-6.pdf}

\bibitem[{Mitre(2012)}]{Mitre}
Mitre, H.~A., 2012. Proposal of game design document from software engineering
  requirements perspective. In: Proceedings of the 2012 17th International
  Conference on Computer Games: AI, Animation, Mobile, Interactive Multimedia,
  Educational \& Serious Games (CGAMES). CGAMES '12. IEEE Computer Society,
  Washington, DC, USA, pp. 81--85.
\newline\urlprefix\url{http://dx.doi.org/10.1109/CGames.2012.6314556}

\bibitem[{Monk and Howard(1998)}]{Monk:1998:MAT:274430.274434}
Monk, A., Howard, S., 1998. Methods \&amp; tools: The rich picture: a tool for
  reasoning about work context. interactions 5~(2), 21--30.

\bibitem[{Natt~och Dag(2002)}]{ElicitMgt}
Natt~och Dag, J., 2002. Elicitation and management of user requirements in
  market-driven software development. Licentiate Thesis.

\bibitem[{Petrillo et~al.(2009)Petrillo, Pimenta, Trindade, and
  Dietrich}]{Petrillo:2009:WWS:1486508.1486521}
Petrillo, F., Pimenta, M., Trindade, F., Dietrich, C., Feb. 2009. What went
  wrong; a survey of problems in game development. Comput. Entertain. 7~(1),
  13:1--13:22.
\newline\urlprefix\url{http://doi.acm.org/10.1145/1486508.1486521}

\bibitem[{Reyno and Cubel(2008)}]{ModelDriven}
Reyno, E.~M., Cubel, J. {\'{A}}.~C., 2008. Model driven game development: 2d
  platform game prototyping. In: GAMEON'2008, (Covers Game Methodology, Game
  Graphics, {AI} Behaviour, Game {AI} Analysis, {AI} Programming, Neural
  Networks and Agent Based Simulation, Team Building, Education and Social
  Networks), November 17-19, 2008, UPV, Valencia, Spain. pp. 5--7.

\bibitem[{Valente and Feij{\'{o}}(2014)}]{ExtendingUseCases}
Valente, L., Feij{\'{o}}, B., 2014. Extending use cases to support activity
  design in pervasive mobile games. In: 2014 Brazilian Symposium on Computer
  Games and Digital Entertainment, {SBGAMES} 2014, Porto Alegre, Brazil,
  November 12-14, 2014. pp. 193--201.
\newline\urlprefix\url{http://dx.doi.org/10.1109/SBGAMES.2014.11}

\bibitem[{Yu(1997)}]{Yu:1997:TMR:827255.827807}
Yu, E. S.~K., 1997. Towards modeling and reasoning support for early-phase
  requirements engineering. In: Proceedings of the 3rd IEEE International
  Symposium on Requirements Engineering. RE '97. IEEE Computer Society,
  Washington, DC, USA, pp. 226--.

\end{thebibliography}

\end{document}